\documentclass[twocolumn,prb,amssymb,amsmath,superscriptaddress,showpac]{revtex4}

\usepackage{mathtext}
\usepackage{amssymb}
\usepackage{graphicx}
\usepackage{amsmath}

\begin{document}

\title{Disorder and temperature renormalization of
interaction contribution to the conductivity in two-dimensional
In$_{x}$Ga$_{1-x}$As electron systems}

\author{G.~M.~Minkov}
\affiliation{Institute of Metal Physics RAS, 620219 Ekaterinburg,
Russia}
\affiliation{Institute of Physics and Applied Mathematics,
Ural State University, 620083 Ekaterinburg, Russia}

\author{A.~V.~Germanenko}
\author{O.~E.~Rut}
\affiliation{Institute of Physics and Applied Mathematics, Ural
State University, 620083 Ekaterinburg, Russia}

\author{A.~A.~Sherstobitov}
\affiliation{Institute of Metal Physics RAS, 620219 Ekaterinburg,
Russia}
\affiliation{Institute of Physics and Applied Mathematics,
Ural State University, 620083 Ekaterinburg, Russia}

\author{B.~N.~Zvonkov}
\affiliation{Physical-Technical Research Institute, University of
Nizhni Novgorod, 603600 Nizhni Novgorod, Russia}

\date{\today}

\begin{abstract}
We study the electron-electron interaction contribution to the
conductivity of  two-dimensional In$_{0.2}$Ga$_{0.8}$As electron
systems in the diffusion regime over the wide conductivity range,
$\sigma\simeq(1-150)\,G_0$, where $G_0=e^2/(2\pi^2\hbar)$. We show
that the data are well described within the framework of the
one-loop approximation of the renormalization group (RG) theory
when the conductivity is relatively high, $\sigma \gtrsim
15\,G_0$. At lower conductivity, the experimental results are
found to be in drastic disagreement with the predictions of this
theory. The theory predicts much stronger renormalization of the
Landau's Fermi liquid amplitude, which controls the interaction in
the triplet channel, than that observed experimentally. A further
contradiction is that the  experimental value of the interaction
contribution does not practically depend on the magnetic field,
whereas the RG theory forecasts its strong decrease due to
decreasing diagonal component of the conductivity tensor in the
growing magnetic field.
\end{abstract}

\maketitle

\section{Introduction}
\label{sec:intr}

A contribution of electron-electron  ({\it e-e}) interaction to
the conductivity is studied since 1980.\cite{Altshuler,Fin} At
high value of the Drude conductivity, $\sigma_0=\pi\, k_Fl\,G_0\gg
G_0$, where $k_F$ is Fermi quasimomentum, $l$ is the mean free
path, and $G_0=e^2/(2\pi^2\hbar)$, and in the diffusion regime,
$T\tau\ll 1$, where $\tau$ is transport relaxation time, this
contribution is:
\begin{eqnarray}
 \delta\sigma_{ee}&=&K_{ee}G_0\ln(T\tau),\nonumber \\
 K_{ee}&=&1+3\left[1-\frac{1+\gamma_2}{\gamma_2}\ln\left(1+\gamma_2\right)\right],
 \label{geq10}
\end{eqnarray}
where $\gamma_2$ stands for the Landau's Fermi liquid amplitude.
The coefficient $K_{ee}$ has two terms coming from singlet and
triplet channels [the first and second terms in Eq.~(\ref{geq10}),
respectively]. They are opposite in sign favoring localization and
antilocalization, respectively. In conventional conductors the
$\gamma _2$ value  is small, and the net effect is in favor of
localization. Together with the weak localization (WL) it leads to
dielectric behavior of the conductivity, $\sigma$: $d\sigma/dT
>0 $. However, the analysis of the {\it e-e} interaction
contribution performed in the framework of the theory of the
renormalization group (RG)\cite{Cast1,Cast2,Cast3,FinRev,Fin1,
Fin2} shows that the reduction of the temperature and/or
conductivity should  lead to renormalization of the Fermi liquid
amplitude $\gamma_2$. At $\sigma_0 \lesssim (5-15)\, G_0 $ or in
dilute systems this amplitude may be significantly enhanced due to
{\it e-e} correlations  that  can result in a metallic-like $T$
dependence of the conductivity: $d\sigma/dT<0$. The theoretical
study within the one-loop approximation for arbitrary valley
degeneracy $n_v$ was carried out in Refs.~\onlinecite{Fin1} and
\onlinecite{Fin2}. The role of two-loop diagrams was studied for
two cases only. The first case relates to multivalley systems
($n_v\gg 1$) with $\gamma_2\ll 1$.\cite{Fin2} The second one is
single valley system ($n_v= 1$) with large $\gamma_2$
value.\cite{Kirkpatr90} The RG theory has been used with advantage
for understanding of the temperature dependence of the
conductivity and metal-insulator transition in
Si-MOSFETs.\cite{Fin1,Knyaz06, Anis07,Knyaz08} As far as we know
there are no data confirming the region of validity of this theory
for the simplest 2D systems with the single valley isotropic
spectrum in the deeply diffusion regime for which the RG equations
were derived.

Besides, the analysis of the temperature dependence of the
conductivity at $B=0$ alone is not reliable way to understand the
role of the renormalization of {\it e-e} interaction and range of
validity of the one-loop approximation. It is because there are
lot of effects, such as the weak localization and
antilocalization, the ballistic contribution of the {\it e-e}
interaction, the temperature dependent screening, the temperature
dependent disorder and so on, which  govern the temperature
dependence of the conductivity along with the {\it e-e}
interaction. Certain of these effects are poorly controlled.
Experimentally, it manifests itself as that the values of the
interaction contribution to the conductivity found from the
temperature dependence of conductivity  at $B=0$ and at $B\neq 0$
are significantly different even in the case of high
conductivity.\cite{Yasin05}

From our point of view the reliable results can be obtained only
from simultaneous analysis of the data obtained at $B=0$ and at
low and high magnetic fields. The unique property of the {\it e-e}
interaction in the diffusion regime is the fact that it
contributes to the diagonal component of the conductivity tensor,
$\sigma_{xx}$, only. Just this feature gives a possibility to
obtain experimentally the {\it e-e} interaction contribution to
the conductivity even for the low conductivity when the
interference contribution dominates.\cite{Minkov03,Minkov07}
Following this line of attack and analyzing the experimental
results obtained for the 2D electron gas in the
GaAs/In$_{0.2}$Ga$_{0.8}$As/GaAs single quantum well the authors
of Ref.~\onlinecite{Minkov03} come to the conclusion that the
temperature dependence of the {\it e-e} interaction contribution
remains logarithmical over the wide conductivity range,
$\sigma_0\simeq(1\ldots 100)\,G_0$: $\delta\sigma_{ee}\simeq
K_{ee}\,G_0\,\ln(T\tau)$. However, the coefficient $K_{ee}$ is
found dependent on the disorder strength. Its value drastically
decreases when $\sigma_0$ decreases, starting from $\sigma_0
\simeq (12-15)\,G_0$. Although this effect is prominent, it was
not since discussed and its origin remained unclear.

In this paper we report the results of the detailed study of the
conductivity of 2D electron gas in In$_{0.2}$Ga$_{0.8}$As and GaAs
single quantum well at $B=0$ and $B\neq 0$ over the wide
conductivity range. We begin by considering the predictions of the
RG theory. Then, after description of experimental details, we
will outline the procedure used for extracting the diffusion part
of the interaction correction. Finally, analyzing the temperature
dependences of the interaction contribution and the conductivity
we will show that the one-loop approximation adequately describes
the data while $\sigma\gtrsim 15\,G_0$ and strongly disagrees with
that at lower conductivity. The conflict between the experiment
and RG theory arising in the presence of the magnetic field will
be discussed as well.

\section{Predictions of the RG theory}

Before to consider and discuss the experimental results let us
demonstrate the role of the $\gamma_2$ renormalization.

\begin{figure}
\includegraphics[width=\linewidth,clip=true]{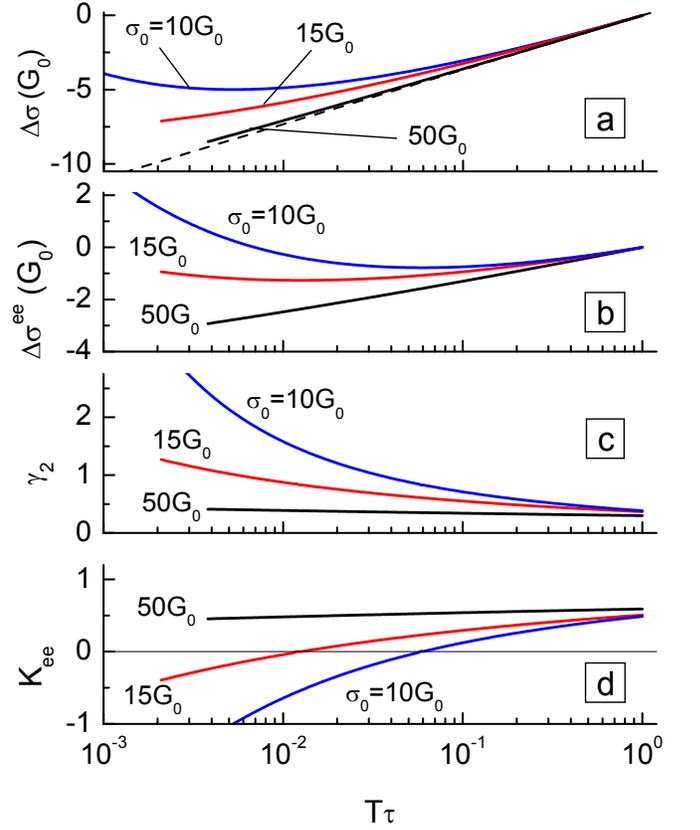}
\caption{(Color online) The temperature dependences of the
conductivity (a), the {\it e-e} interaction contribution to the conductivity (b), the Fermi liquid  amplitude
$\gamma_2$ (c), and  $K_{ee}$ (d) found from the solution of
Eqs.~(\ref{eq5}) and (\ref{eq6}).  The dashed
line is the dependence $1.59 G_0\ln(T\tau)$. The following parameters have been used:
$n=2.0\times 10^{11}$~cm$^{-2}$, $2.5\times 10^{11}$~cm$^{-2}$, and $5.0\times 10^{11}$~cm$^{-2}$ and $\gamma_2^0=0.383$, $0.367$, and $0.3$
for $\sigma_0=10\,G_0$, $15\,G_0$, and  $50\,G_0$, respectively.
}\label{F1}
\end{figure}

The temperature dependence of the conductivity $\sigma$ and the
Fermi liquid  amplitude $\gamma_2$ is described in the
framework of one-loop approximation of RG theory by the
following system of the differential
equations:\cite{Fin2,FinRev,Cast1,Cast2,Cast3,Fin1}
\begin{eqnarray}
\frac{d\sigma}{d\xi}& =&
-\left\{1+1+3\left[1-\frac{1+\gamma_2}{\gamma_2}\ln(1+\gamma_2)\right]\right\}
\label{eq5}\\
\frac{d\gamma_2}{d\xi}&=&\frac{1}{\sigma}\frac{\left(1+\gamma_2\right)^2}{2}
\label{eq6}
\end{eqnarray}
where $\xi=-\ln(T\tau)$, and $\sigma$ is measured in units of
$G_0$. The quantity $\gamma_2$ is expressed through the Fermi
liquid constant $F_0^\sigma$:
$\gamma_2=-F_0^\sigma/\left(1+F_0^\sigma\right)$. For the high
conductivity, the value of $F_0^\sigma$ depends on the gas
parameters $r_s=\sqrt{2}/(a_Bk_F)$, where $a_B$ is the effective
Bohr radius, and  for small $r_s$ values  is\cite{Aleiner1}
\begin{equation}
 F_0^\sigma =-\frac{1}{2\pi}\frac{r_s}{\sqrt{2-r_s^2}}
  \ln\left(\frac{\sqrt{2}+\sqrt{2-r_s^2}}{\sqrt{2}-\sqrt{2-r_s^2}}\right),\,\,\, r_s^2<2.
 \label{F0s}
\end{equation}
The term $1+1$ in braces in Eq.~(\ref{eq5}) is responsible for
the weak localization and the interaction in singlet channel,
which in the case of Coulomb interaction give equal
contributions. Eq.~(\ref{eq6}) describes the renormalization of
the Landau's Fermi liquid amplitude $\gamma_2$ with the
temperature and conductivity. One can see from Eq.~(\ref{eq6})
that the $\gamma_2$ renormalization can be neglected at high
conductivity. In this case the integration of Eq.~(\ref{eq5})
gives
\begin{equation}
\sigma(T)= \left(1+K_{ee}\right)G_0\ln(T\tau)+\text{const}
 \label{AA}
\end{equation}
with $K_{ee}$ given by Eq.~(\ref{geq10}). This expression accords
well with the known expression
\begin{equation}
\sigma(T)=\sigma_0+K_{ee}G_0\ln(T\tau)+G_0\ln{\left[\frac{\tau}{\tau_\phi(T)}\right]},
 \label{EqSigma}
\end{equation}
where $\tau_\phi(T)\propto 1/T$ is the phase relaxation time
controlled in 2D systems  by the inelasticity of the {\it e-e}
interaction.  However, Eqs.~(\ref{eq5}) and (\ref{eq6}) predict
that the change of the amplitude $\gamma_2$ and, hence, the
deviation of the temperature dependence of the conductivity from
the logarithmic one is appreciable already at moderate
conductivity value. The written is illustrated by Fig.~\ref{F1},
in which the results of numerical solution of Eqs.~(\ref{eq5}) and
(\ref{eq6}) are presented. We used the parameters, which are
typical for the moderately disordered
GaAs/In$_{0.2}$Ga$_{0.8}$As/GaAs heterostructures investigated in
this paper. The minimal $T\tau$ value corresponds to $T=0.1$~K for
all the cases. The following initial conditions have been used. We
suppose that the high-temperature conductivity is equal to the
Drude conductivity: $\sigma(\xi=0)=\sigma_0$. This condition seems
to be natural. It corresponds to that the diffusion part of
interaction correction is equal to zero and the WL correction is
much less than the Drude conductivity at $T\tau=1$. The second
condition is $\gamma_2(\xi=0)\equiv
\gamma_2^0=-F_0^\sigma/\left(1+F_0^\sigma\right)$, where
$F_0^\sigma$ is determined by Eq.~(\ref{F0s}).

One can see from Fig.~\ref{F1}(c) that the renormalization of
$\gamma_2$ for the high Drude conductivity, $\sigma_0=50\,G_0$, is
rather small so that the temperature dependence of the
conductivity is close to the logarithmic one with the slope
determined by the initial value of $\gamma_2$:
$\Delta\sigma(T)=\left[1+K_{ee}(\gamma_2^0)\right]G_0\ln{T\tau}=1.59G_0\ln{T\tau}$
[Fig.~\ref{F1}(a)]. Nevertheless, the noticeable decrease of the
$K_{ee}$ value with the lowering temperature is evident even for
so high conductivity [see Fig.~\ref{F1}(d)]. The $K_{ee}$ value at
$T\tau= 3.5\times 10^{-3}$ is approximately equal to $0.46$, while
$K_{ee}$ at $T\tau=1$ is close to  $0.6$.

For the lower Drude conductivity, $\sigma_0=15\,G_0$, the
renormalization of $\gamma_2$ with the temperature decrease
becomes significant [Fig.~\ref{F1}(c)]. The sign of
$d\Delta\sigma^{ee}/dT$ is changed at $T\tau\simeq 0.012$ from
positive at high temperature to negative at lower one
[Fig.~\ref{F1}(b)].  However, the temperature dependence of the
overall conductivity remains insulating ($d\sigma/dT>0$) due to
dominating WL contribution. Finally, for $\sigma_0=10\,G_0$, the
renormalization of $\gamma_2$ is so huge [Fig.~\ref{F1}(c)] that
the metallic behavior of the interaction correction
[Fig.~\ref{F1}(b)] wins the insulating behavior of the WL
correction at low temperature, and, as consequence, the total
conductivity behaves itself metallically at
$T\tau\lesssim5\times10^{-3}$ [Fig.~\ref{F1}(a)]. To the best of
our knowledge such the behavior was never experimentally observed
in the moderately disordered 2D systems of weakly interacting
electrons with the simple single-valley energy spectrum,
characterizing by $r_s<2-3$ and $\sigma\gtrsim 1\,G_0$. The goal
of this paper is to examine how the one-loop approximation
describes the experimental data for such the systems and, thus,
establish the region of validity of this theory.

\section{Experiment}

The results of experimental study of the evolution of the
diffusion part of the interaction correction to the conductivity
in a $n$-type 2D system with decreasing Drude conductivity within
the range from $\sigma_0\simeq 150\,G_0$ to $\sigma_0\simeq
5\,G_0$ at the temperatures when $T\tau<0.1-0.15$ are reported.
The ballistic contribution of the {\it e-e} interaction is small
under these conditions. The data for two structures, 3510 and
4261, are analyzed. The structure 3510 with moderate disorder has
two $\delta$ doping layers disposed in the barriers on each side
of the quantum well on the distance of about $9$~nm. The structure
4261 with higher disorder has the $\delta$ layer in the center of
the quantum well. In more detail the structures design is
described in Refs.~\onlinecite{our5} and \onlinecite{Minkov07-1}.
The electron density $n$ and mobility $\mu$ in the structures are
as follows: $7.0\times 10^{11}$, $\mu=19300$~cm$^2$/V~s for
structure 3510, and $n=1.8\times 10^{12}$~cm~$^{-2}$ and
$\mu=1600$~cm$^2$/V~s for structure 4261. The samples were mesa
etched into standard Hall bars and then an Al gate was deposited
by thermal evaporation onto the cap through a mask. Varying the
gate voltage, we changed the electron density in the quantum well
and changed the conductivity from its maximal value down to
$\sigma\simeq 1\,G_0$.

Firstly, let us demonstrate that the structures investigated are
``normal'', i.e., the transport in zero, low and high magnetic
field at the high conductivity, when the renormalization of the
{\it e-e} interaction should be negligible, is consistent with the
following simple model. The temperature dependence of the
conductivity in the absence of magnetic field can be described by
Eq.~(\ref{EqSigma}), whereas in the presence of the magnetic field
the conductivity tensor components are
\begin{eqnarray}
\sigma_{xx}(B,T)&=&\frac{en\mu(B,T)}{1+\left[\mu(B,T)B\right]^2}+\delta\sigma^{ee}(T) \label{sxx} \\
\sigma_{xy}(B,T)&=&\frac{en\mu(B,T)^2B}{1+\left[\mu(B,T)B\right]^2}.\label{sxy}
\end{eqnarray}
Because the WL correction is actually reduced to the
renormalization of the transport relaxation time,\cite{Dmitriev97}
it is incorporated here into the mobility in such a way that
\begin{equation}
\delta\sigma^{WL}(B,T)=e\,n\,\delta\mu(B,T),
\label{EqMu}
\end{equation}
where
\begin{equation}
\delta\sigma^{WL}(B=0,T)=-G_0 \ln{\left[\frac{\tau}{\tau_\phi(T)}\right]},
\label{EqWL}
\end{equation}
and
$\Delta\sigma^{WL}(B)=\delta\sigma^{WL}(B)-\delta\sigma^{WL}(B=0)$
is described by the expression\cite{Hikami80,Wittman87}
\begin{eqnarray}
\Delta\sigma^{WL}(B)=\alpha\,G_0{\cal H}\left(\frac{\tau}{\tau_\phi},\frac{B}{B_{tr}}\right), \nonumber \\
{\cal H}(x,y)=\psi\left(\frac{1}{2}+\frac{x}{y}\right)-\psi\left(\frac{1}{2}+\frac{1}{y}\right)-\ln{x}.
\label{HLN}
\end{eqnarray}
Here, $B_{tr}=\hbar/(2el^2)$ is the transport magnetic field,
$\psi(x)$ is a digamma function, and $\alpha$ is the prefactor,
which value depends on the conductivity if one takes into account
two-loop localization correction and the interplay of the weak
localization and interaction:
$\alpha=1-2G_0/\sigma$.\cite{Aleiner99,Minkov04}

\begin{figure}
\includegraphics[width=\linewidth,clip=true]{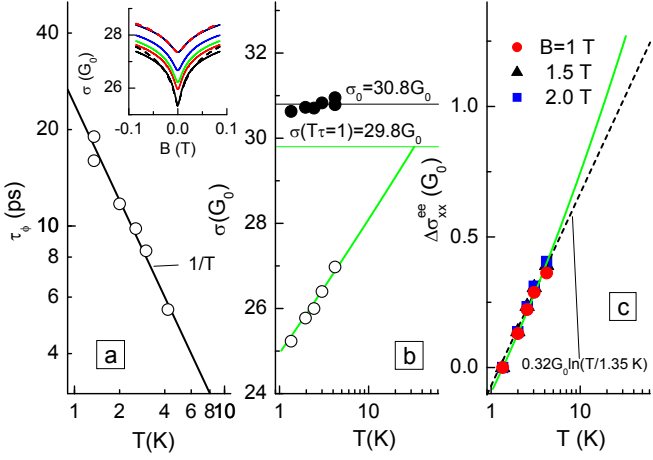}
\caption{(Color online)  The temperature dependences of the
phase relaxation time (a), $\sigma$ and $\sigma_0$ found from
experiment (see text) (b),  and $\Delta\sigma_{xx}^{ee}$
taken at different magnetic field (c).  Inset in panel (a) shows the $\sigma$~vs~$B$
dependence for $T=4.2,\, 3.0,\, 2.56,\, 2.0,\, 1.35$~K (from top to bottom). The dashed lines  are the
results of the best fit by Eq.~(\ref{HLN}) carried out at $|B|<0.3B_{tr}$ ($B_{tr}=80$ mT for this case.)
Solid lines in (b) and (c) are the solutions of the RG equations
with the initial conditions: $\sigma(T\tau=1)=29.8\,G_0$, $\gamma_2^0=0.4$. Structure 3510,
$n=3.35\times 10^{11}$~cm$^{-2}$.}
\label{F2}
\end{figure}

For structure 3510, the low-field magnetoconductivity
$1/\rho_{xx}(B)$, which results from the suppression of the
interference quantum correction, measured at high conductivity for
different temperatures is presented in inset in Fig.~\ref{F2}(a).
One can see that the data are well described by Eq.~(\ref{HLN}).
The temperature dependence of $\tau_\phi$ within the experimental
accuracy is close to $1/T$ [Fig.~\ref{F2}(a)]. This shows that the
main mechanism of the phase relaxation is, as expected,
inelasticity of the {\it e-e} scattering. The prefactor value is
close to unity.

To find the diffusion part of the interaction correction  we take
approach which has been detailed in our previous paper,
Ref.~\onlinecite{our5}. It uses the unique property of the
diffusion correction to contribute to $\sigma_{xx}$ and do not to
$\sigma_{xy}$, see Eqs.~(\ref{sxx}) and (\ref{sxy}). Thus, in
order to obtain the correction experimentally one should find such
the contribution to the conductivity which exists in $\sigma_{xx}$
but is absent in $\sigma_{xy}$. The temperature dependences of
$\Delta\sigma_{xx}^{ee}=\delta\sigma_{xx}^{ee}(T)-\delta\sigma_{xx}^{ee}(1.35\text{~K})$
found in such the way for the different magnetic fields are shown
in Fig.~\ref{F2}(c). One can see that these dependences are
logarithmic within the experimental accuracy,
$\Delta\sigma_{xx}^{ee}=K_{ee}G_0 \ln(T/1.35\text{~K})$, and
$K_{ee}\simeq 0.32$  does not depend on the magnetic field. The
temperature dependence of the conductivity at $B=0$ is shown in
Fig.~\ref{F2}(b). As seen it is also logarithmic, and, what is
more important, the slope of the $\sigma$~vs~$\ln{T}$ dependence
is close to the value $1+K_{ee}=1.32$ predicted theoretically for
the case when only the WL and interaction correction are
responsible for the temperature dependence of $\sigma$ [see
Eq.~(\ref{AA})]. This fact justifies that there are no additional
mechanisms of the $T$-dependence of the conductivity in the
samples investigated. It is wholly determined by the temperature
dependence of the WL and interaction quantum corrections. Now,
knowing the experimental $K_{ee}$ and $\tau_\phi$ values one can
easily estimate the value of the Drude conductivity with the use
of Eq.~(\ref{EqSigma}). As seen from Fig.~\ref{F2}(b) the values
of $\sigma_0$ found at different temperatures are very close to
each other. This attests that the model is adequate and the value
of $\sigma_0$ found in this way is a good estimate for the Drude
conductivity. Thus, $\sigma_0=(30.8\pm 0.2)\,G_0$ for this case.

Let us compare the experimental temperature dependences of
conductivity with that predicted by the RG theory. Solid line in
Fig.~\ref{F2}(b) is the result of the numerical solution of
Eqs.~(\ref{eq5}) and (\ref{eq6}) with the initial parameters which
give the best fit of the data: $\sigma(T\tau=1)=29.8\,G_0$ and
$\gamma_2^0=0.4$. The variation of the {\it e-e} interaction
contribution  shown in Fig.~\ref{F2}(c) has been obtained by
subtraction of the WL contribution from the calculated
$\sigma$~vs~$T$ curve as follows:
$\Delta\sigma^{ee}(T)=\sigma(T)-\sigma(1.35\text{
K})-\ln(T/1.35\text{ K})$. Excellent agreement between the data
and solution of the RG equations  is evident both for $\sigma(T)$
and for $\Delta\sigma_{xx}^{ee}(T)$.

It is worth noting that the values of $\sigma(T\tau=1)$ and
$\gamma_2^0$ found from the fit are reasonable. The initial value
of $\gamma_2$, $\gamma_2^0=0.42$, is  close to that calculated
from Eq.~(\ref{F0s}), $\gamma_2=0.37$. The value of $\sigma$ at
$T\tau=1$ is less than the Drude conductivity estimated
experimentally by the value of about $1\,G_0$. The reason is that
not all the interference quantum correction is suppressed at
$T\tau=1$. Really, extrapolating the experimental
$\tau_\phi$~vs~$T$ dependence to $T=\tau^{-1}\simeq 30$~K we
obtain for the rest of $\delta\sigma^{WL}$ at $T\tau=1$:
$\delta\sigma^{WL}(30~\text{K})\simeq
G_0\ln{\left[\tau/\tau_\phi(30~\text{K})\right]}\simeq -1.2\,G_0$.
Thus, the Drude conductivity estimated as
$\sigma(30~\text{K})-\delta\sigma^{WL}(30~\text{K})$ is
$\sigma_0\simeq (29.8+1.2)\,G_0=31\,G_0$. This value practically
coincides with that obtained above, $\sigma_0=(30.8\pm 0.2)\,G_0$.

It  would be fine to trace experimentally the $\gamma_2$ change
over the whole temperature range starting from $T\tau=1$. However,
the ballistic contribution of the interaction correction, the
partial lifting of the degeneracy of the electron gas, finally,
the phonon scattering control the temperature dependence of the
conductivity at high temperature. All this makes it impossible to
determine the {\it e-e} interaction contribution accurately
already at $T\tau\gtrsim 0.1-0.15$. On the other hand,
Fig.~\ref{F1} shows that the renormalization of $\gamma_2$
strongly depends on the value of the Drude conductivity, it should
be more pronounced at lower Drude conductivity.
\begin{figure}
\includegraphics[width=0.8\linewidth,clip=true]{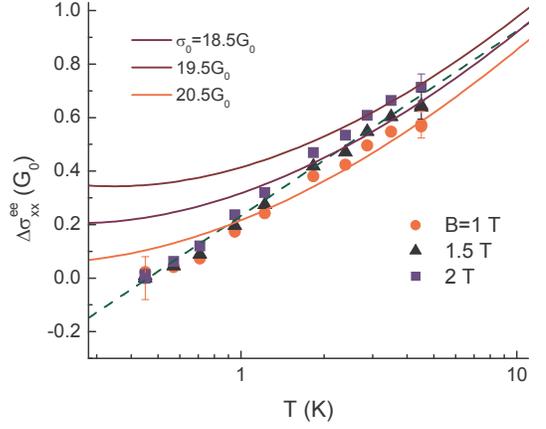}
\caption{(Color online) The temperature dependences of
$\Delta\sigma_{xx}^{ee}$
for different magnetic fields for structure 3510 at $\sigma_0=18.5\, G_0$, $n=3\times 10^{11}$~cm$^{-2}$.
The curves
are the solutions of the RG equations for different $\sigma_0$
and $\gamma_2^0=0.41$. For clarity the curves are shifted in vertical direction.
The dashed line is the dependence $0.3G_0\ln(T/0.45\text{~K})$. }
\label{dSxsDr18}
\end{figure}

The  analysis described above  has been carried out over the wide
range of the gate voltage which controls the electron density, the
mobility and, thus, the Drude conductivity. All the dependences,
namely $\sigma(T)$ at $B=0$, $\tau_\phi(T)$, and $\sigma_{xx}(T)$
are similar to that shown in Fig.~\ref{F2}. However, agreement of
the data with the solution of Eqs.~(\ref{eq5}) and (\ref{eq6}) is
the worse, the lower the conductivity.

Disagreement becomes noticeable already at $\sigma_0\simeq
19\,G_0$. It is more visible in $\Delta\sigma_{xx}^{ee}$~vs~$T$
dependence (see Fig.~\ref{dSxsDr18}). The experimental dependence
remains close to the logarithmic one, while the curve calculated
with the initial value
$\sigma(T\tau=1)=\sigma_0+\delta\sigma^{WL}(T\tau=1)=18.5\,G_0$
shows upturn at $T\lesssim 0.3$~K. The $\gamma_2$ value found from
the slope of the experimental dependence is approximately equal to
$0.55$, whereas the calculated value of $\gamma_2$ changes from
$\gamma_2=0.57$ at $T=4.2$ K to $\gamma_2=0.8$ at $T=0.45$~K.
Variation of the initial conditions within the reasonable range
does not improve agreement. The dependence calculated remains
nonlogarithmic.

\begin{figure}
\includegraphics[width=\linewidth,clip=true]{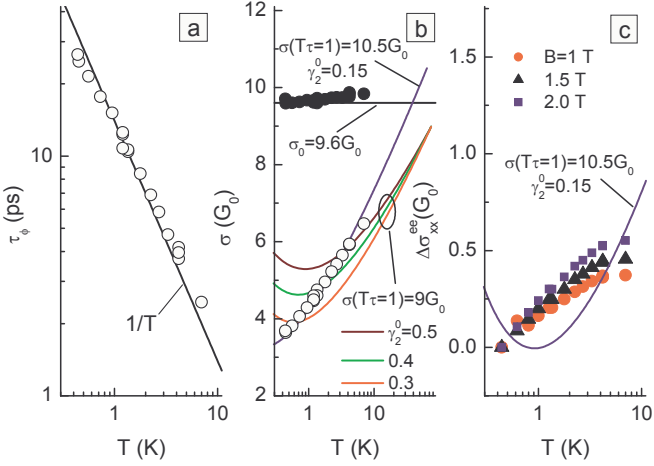}
\caption{(Color online) The temperature dependences of the
phase relaxation time (a), $\sigma$ and $\sigma_0$ found from
the experiment (b), and $\Delta\sigma_{xx}^{ee}$
taken at different magnetic field (c) for the structure 3510 at
$\sigma_0=9.6\,G_0$, $n=2.3\times 10^{11}$~cm$^{-2}$.  The curves in panels
(b) and (c) are the solutions of the RG equations, Eqs.~(\ref{eq5}) and (\ref{eq6}),
with different initial conditions.}
\label{F3}
\end{figure}

The distinction between the calculation and experimental results
becomes more clear at lower $\sigma_0$. As an example we present
the data for $\sigma_0\simeq 9.6\, G_0$ in Fig.~\ref{F3}. It is
seen from Fig.~\ref{F3}(b) that the temperature dependence of the
conductivity remains close to the logarithmic one. The temperature
dependence of $\Delta\sigma_{xx}^{ee}$ is also close to the
logarithmic one with the slope corresponding to
$\gamma_2=0.64-0.68$.

As in the case of the higher conductivity, the temperature
dependence of $\tau_\phi$ is close to $1/T$ [Fig.~\ref{F3}(a)]
therefore the term responsible for the weak localization in
Eq.~(\ref{eq5}) remains equal to $1$. However, it is impossible to
describe the temperature dependence of the conductivity for $B=0$
if one uses $\sigma(T\tau=1)$ found from the experimental Drude
conductivity as in the previous case [Fig.~\ref{F3}(b)]. One can
suppose that the value of $\sigma(T\tau=1)$ has been obtained with
large error and another value should be used as initial one. We
tried to describe the data using both $\sigma(T\tau=1)$ and
$\gamma_2^0$ as the fitting parameters. As seen from
Fig.~\ref{F3}(b) the much better agreement can be achieved in this
case. However, as clearly evident from Fig.~\ref{F3}(c), even with
these parameters the calculated dependence
$\Delta\sigma^{ee}_{xx}(T)$ strongly deviates from the
experimental one. Namely, the upturn of $\Delta\sigma_{xx}^{ee}$
with the temperature decrease predicted by the RG theory is not
observed experimentally.
\begin{figure}
\includegraphics[width=\linewidth,clip=true]{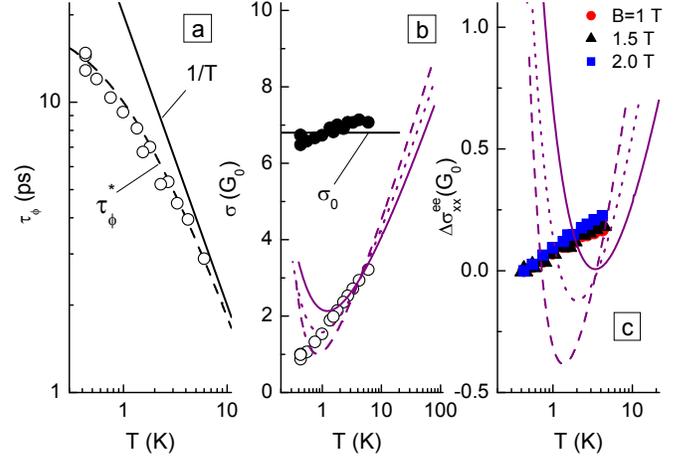}
\caption{(Color online) The temperature dependences of the
phase relaxation time (a), $\sigma$ and $\sigma_0$ found from
experiment (b), and $\Delta\sigma_{xx}^{ee}$
taken at different magnetic field (c) for structure 3510 at $\sigma_0=6.8\,G_0$, $n=2\times 10^{11}$~cm$^{-2}$.
The dashed line in (a) is the dependence $\tau_\phi^\star(T)$ with
$\tau_\phi=20/T$,~ps and $\xi=15\,l$.
The lines in (b) and (c) are the solutions of the RG equations
with parameters $\sigma(T\tau=1)=7.5\,G_0$, $\gamma_2^0=0.12$
(solid lines), $\sigma(T\tau=1)=8\,G_0$, $\gamma_2^0=0.01$ (dotted
lines), and $\sigma(T\tau=1)=8.5\,G_0$, $\gamma_2^0=-0.1$ (dashed
lines).} \label{F4}
\end{figure}

It could be  assumed that the procedure of the extraction of  the
{\it e-e} interaction contribution being transparent nevertheless
fails. However, the RG theory predicts that not only the {\it e-e}
interaction contribution $\delta\sigma^{ee}$ should demonstrate
the upturn with the temperature decrease at the lower Drude
conductivity but the total conductivity $\sigma$ as well [see
Fig.~\ref{F1}(a)]. In this case the experiment and RG theory can
be compared directly without any additional treatment of the data.
Therefore, let us inspect the results for the lower Drude
conductivity, $\sigma_0\simeq 6.8\, G_0$, presented in
Fig.~\ref{F4}. The temperature dependence of $\tau_\phi$ found
from the low-field negative magnetoresistance deviates from the
$1/T$ law demonstrating tendency to saturation at low temperature
[see Fig.~\ref{F4}(a)]. As shown in Ref.~\onlinecite{Minkov07-1}
such the behavior results from the fact that the dephasing length
$L_\phi=\sqrt{D\tau_\phi}$ (where $D$ is the diffusion
coefficient) at low $T$ becomes comparable with the localization
length $\xi\sim l \exp{\left(\pi k_Fl/2\right)}$, and the quantity
$\tau_\phi^\star=1/\left(1/\tau_\phi+D/\xi^2\right)$, rather than
$\tau_\phi$ is experimentally obtained from the fit of the
magnetoresistance. Indeed, the data in Fig.~\ref{F4}(a) are well
described by this formula with $\tau_\phi=20/T$,~ps  and
$\xi=15\,l$ that is close to $\exp{\left(\pi k_Fl/2\right)}\simeq
30\,l$. Since the $\tau_\phi$ saturation is not yet observed in
our temperature range, the temperature dependence of the
conductivity remains close to the logarithmic one [see open
symbols in Fig.~\ref{F4}(b)]. In contrast to that, the RG
equations predict the upturn of the conductivity within our
temperature range independently of the initial $\sigma$ and
$\gamma_2$ values: $\sigma(T\tau=1)$ and $\gamma_2^0$. As
discussed above the upturn results from the strong renormalization
of $\gamma_2$ that leads not only to the change of sign of
$d\sigma^{ee}/d T$, but to large its value as well, so that
$|d\sigma^{ee}/d\ln{T}|>d\sigma^{WL}/d\ln{T}$. In the actual case
this results in that the calculated curve following the data at
the hight temperature, $T\simeq 1.5-6$~K, exhibits, nevertheless,
minimum at $T\simeq 1.2$~K and growth at lower temperature [see
dotted curve in Fig.~\ref{F4}(b)].

\begin{figure}
\includegraphics[width=0.75\linewidth,clip=true]{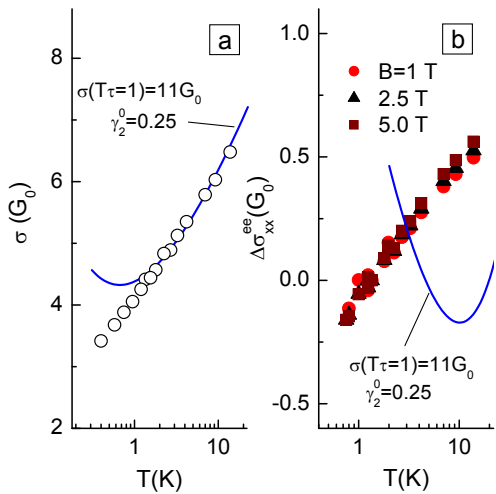}
\caption{(Color online) The temperature dependences of the
$\sigma$ (a) and $\Delta\sigma_{xx}^{ee}$
taken at different magnetic field (b) for the structure 4261 at $\sigma_0=10.2\,G_0$ and  $n=1.05\times 10^{12}$~cm$^{-2}$.
The curves are the solutions of the RG equations, Eqs.~(\ref{eq5}) and (\ref{eq6}),
with parameters corresponding to the best fit of the $\sigma$~vs~$T$ data.}
\label{4261}
\end{figure}

It is clear that the one-loop approximation of the RG theory is
insufficient for so low conductivity, $\sigma\sim 1\,G_0$. It is
pertinent to note here, that such decisive disagreement with the
RG theory for the structure 4261 with the stronger disorder is
observed at the higher conductivity (Fig.~\ref{4261}). As seen the
experimental $T$ dependence of $\sigma$ is close to the
logarithmic one,\cite{fnt2} whereas the RG equations predict the
upturn in the $\sigma$~vs~$T$ plot already at $\sigma\simeq
4.5\,G_0$.

\begin{figure}
\includegraphics[width=0.8\linewidth,clip=true]{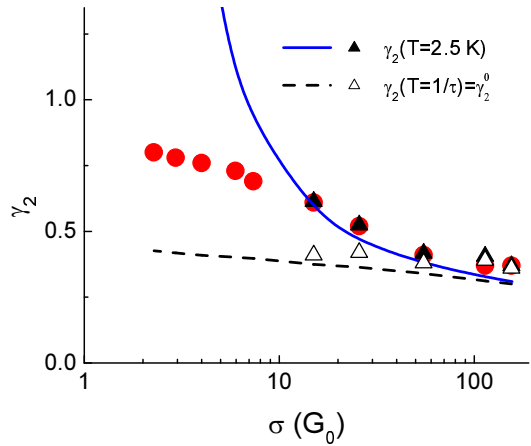}
\caption{(Color online) The value of  $\gamma_2$ found experimentally (symbols) and
calculated from the RG equations for $T=2.5$~K (solid line) as a function of the  conductivity at $T=4.2$~K for
the sample 3510.
The dashed line is the $\sigma$ dependence of $\gamma_2^0$ calculated with the help of
Eq.~(\ref{F0s}). Triangles correspond to the best fit of the data by Eqs.~(\ref{eq5}) and (\ref{eq6}).
Circles are
obtained with the help of Eq.~(\ref{geq10}) from the slope of the
experimental $\Delta\sigma_{xx}^{ee}$~vs~$\ln{T}$ dependence
within the temperature range from $1.3$~K to $4.2$~K.}
\label{F5}
\end{figure}

Let us analyze the  results in the whole. In Fig.~\ref{F5} we
compare the low temperature values of the Fermi liquid amplitude
$\gamma_2$ obtained experimentally in wide range of the
conductivity values, $\sigma=(2-150)\,G_0$,\cite{fnt1} with that
predicted by the RG theory. The $\gamma_2$ data obtained with the
help of Eq.~(\ref{geq10}) from the slope of the experimental
$\Delta\sigma_{xx}^{ee}$~vs~$\ln{T}$ dependence within the
temperature range from $1.3$~K to $4.2$~K are shown by circles.
They agree well with the results obtained from the fit of the
$\Delta\sigma_{xx}^{ee}$~vs~$T$ data by the RG equations,
Eqs.~(\ref{eq5}) and (\ref{eq6}) (shown by solid triangles). As
shown above such the fit is possible only at the relatively high
conductivity, $\sigma_0\gtrsim 18\,G_0$. Note that the initial
values of $\gamma_2$, $\gamma_2^0$, fall closely to the curve
calculated from Eq.(\ref{F0s}). One can see that $\gamma_2$
increases monotonically when the conductivity goes down, and  this
increase is well described by the one-loop RG equations down to
$\sigma\simeq 15\,G_0$. At the lower conductivity, the theory
predicts much steeper rise of $\gamma_2$ than that obtained from
the data treatment.

\begin{figure}
\includegraphics[width=0.8\linewidth,clip=true]{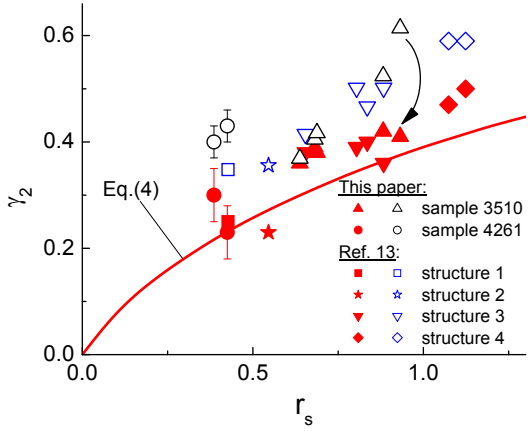}
\caption{(Color online) The Landau's Fermi liquid amplitude $\gamma_2$ plotted against the gas
parameter $r_s$ for the samples 3510 and 4261, and for the structures from Ref.~\onlinecite{Minkov03}.
Open symbols are obtained from the experimental value of
$K_{ee}$ at low temperature, $T=(1.3-4.2)$~K. Solid symbols are the values of $\gamma_2^0$
corresponding to the best fit of the experimental $\Delta\sigma_{xx}^{ee}$~vs~$T$ dependence by
the solution of the RG equations. Arrow indicates the renormalization of $\gamma_2$ with
the lowering temperature.
Only the data for $\sigma> 15\,G_0$ are presented.} \label{F6}
\end{figure}

As mentioned in Sec.~\ref{sec:intr} the behavior of the {\it e-e}
interaction contribution with decreasing conductance was studied
in Ref.~\onlinecite{Minkov03}. It has been, in particular, found
that the value of $K_{ee}$ decreases with the $\sigma_0$ decrease.
We have retreated those data following the line of attack
described above. The results for $\sigma>15\,G_0$ are presented in
Fig.~\ref{F6}. In this figure, the open symbols are the low
temperature $\gamma_2$ value obtained from
$\Delta\sigma_{xx}^{ee}$~vs~$T$ dependence. The value of
$\gamma_2^0$ corresponding to the best fit of the data by the RG
equations are presented by solid symbols. It is seen that while
the low temperature $\gamma_2$ points lie noticeably above the
theoretical curve, Eq.~(\ref{F0s}), the $\gamma_2^0$ values fall
very close to it  for all the structures investigated both in
Ref.~\onlinecite{Minkov03} and in this paper.

Thus, the RG equations well describe both the temperature
dependence of the conductivity at $B=0$ and the temperature
dependence of the interaction contribution at the relatively high
conductivity, $\sigma\gtrsim 15\,G_0$. The renormalization of the
interaction constant $\gamma_2$ at $\sigma=15\,G_0$ is large
enough. For instance, the amplitude $\gamma_2$ in the sample 3510
increases from $0.41$ at $T=50$~K up to $0.61$ at $T=2.5$~K. Such
the increase corresponds to the reduction of $K_{ee}$ by a factor
of two: from $0.46$ at the higher temperature down to  $0.23$ at
the lower one. The renormalization of $\gamma_2$ becomes stronger
at further conductivity decrease. However, the experimental
$\gamma_2$~vs~$\sigma$ plot saturates when $\gamma_2$ increases by
about of two times in our conductivity range, whereas the RG
theory predicts much stronger renormalization at $\sigma<10\,G_0$
(see Fig.~\ref{F5}). One of the possible reason of the
disagreement between the theory and experiment at $\sigma<15\,G_0$
is the restriction of the one-loop approximation.  Another
possible reason is the interplay of the weak localization and
interaction, which is not taken into account in the theory
considered.

It would seem that the understanding is achieved: the RG theory
adequately describes  the data while the conductivity is rather
high, $\sigma\gtrsim 15\,G_0$.

\begin{figure}
\includegraphics[width=\linewidth,clip=true]{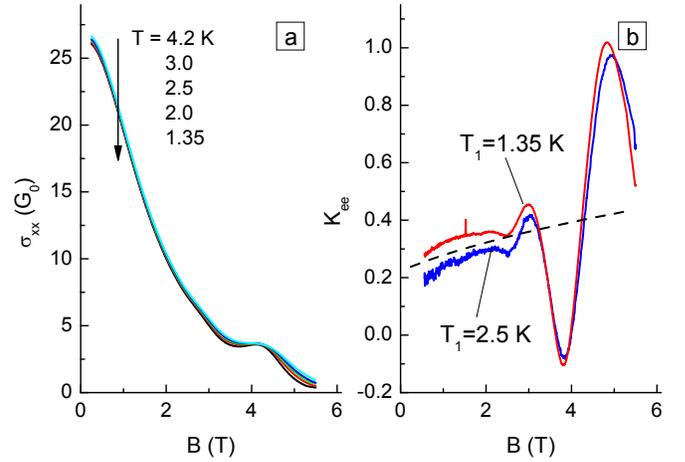}
\caption{(Color online) The magnetic field dependence of $\sigma_{xx}$ (a) and $K_{ee}=\left[\sigma_{xx}^{ee}
(T_1)-\sigma_{xx}^{ee}(4.2\text{~K})\right]/G_0\ln{(T_1/4.2\text{ K})}$ (b)
for the structure 3510 at $\sigma_0\simeq 30.8\,G_0$. The dashed line indicates
the monotonic run of the $K_{ee}$~vs~$B$ curves.}
\label{F8}
\end{figure}

However, let us analyze the data in the presence of  magnetic
field in more detail.  Eqs. (\ref{eq5}) and (\ref{eq6}) have to
describe $\sigma_{xx}(T)$ if the  Zeeman splitting is relatively
small, $|\textsl{g}|\mu_B B<T$. In the case of $B>B_{tr}$, when
the WL contribution is suppressed, the one unity in braces in
Eq.~(\ref{eq5}) should be omitted. Because $\sigma_{xx}$ strongly
decreases with increasing $B$, $\sigma_{xx}\propto 1/B^2$,  when
$B>1/\mu$, the strong change of the interaction contribution due
to the renormalization of $\gamma_2$ should be clearly evident
also. As evident from Figs.~\ref{F2}(c), \ref{dSxsDr18},
\ref{F3}(c), \ref{F4}(c), and \ref{4261}(b), the
$\Delta\sigma_{xx}^{ee}$~vs~$T$ plots are practically independent
of the magnetic field. This contradiction is more clearly
illustrated by Fig.~\ref{F8}, where the results for structure 3510
at $\sigma_0\simeq 30.8\,G_0$ are presented. As seen from
Fig.~\ref{F8}(a)  $\sigma_{xx}$  decreases drastically: from
$\simeq 25\,G_0$ at $B=0.5$~T down to $\simeq 2\,G_0$ at $B=5$~T.
However, the quantity
$\left[\delta\sigma_{xx}^{ee}(T_1)-\delta\sigma_{xx}^{ee}(4.2\text{~K})\right]/G_0\ln{(T_1/4.2\text{
K})}$, which is $K_{ee}$ according to Eq.~(\ref{geq10}), slightly
increases against the background of the Shubnikov-de Haas  (SdH)
oscillations [Fig.~\ref{F8}(b)], instead of decreasing due to the
renormalization of $\gamma_2$. It would be assumed that the
decrease expected is compensated by the increase of
$\Delta_{xx}^{ee}$ caused by the suppression of two from three
triplet channels due to the Zeeman
effect.\cite{Cast1,Cast3,Fin84,Raim90,Minkov07} However, this
effect becomes essential in strong magnetic fields,
$|\textsl{g}|\mu_BB>T$, which is not the case.

\begin{figure}
\includegraphics[width=\linewidth,clip=true]{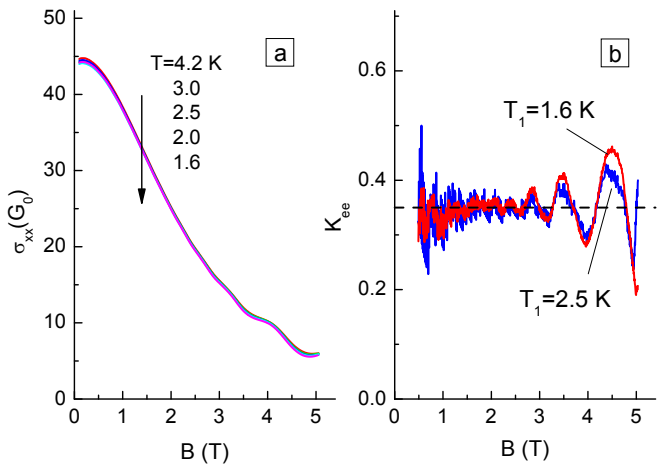}
\caption{(Color online) The magnetic field dependence of $\sigma_{xx}$ (a) and $K_{ee}$ (b)
for the structure Al$_{0.3}$Ga$_{0.7}$As/GaAs/Al$_{0.3}$Ga$_{0.7}$As
with zero $\textsl{g}$-factor at $\sigma_0\simeq 47\,G_0$, $n=8\times10^{11}$~cm$^{-2}$.
The dashed line is provided as a guide for
the eye, it indicates an absence of the monotonic component in $K_{ee}$~vs~$B$ dependences.}
\label{F7}
\end{figure}

To assure that the {\it e-e} interaction  contribution does not
depend on the magnetic field in spite of strong $\sigma_{xx}$
decrease, we have performed analogous measurements on the
structure with the very small value of  $\textsl{g}$-factor. It is
the Al$_{0.3}$Ga$_{0.7}$As/GaAs/Al$_{0.3}$Ga$_{0.7}$As structure
with the quantum well width $8$~nm. The electron density in the
structure is $n=8\times10^{11}$~cm$^{-2}$. The value of
$\textsl{g}$-factor on the bottom of the conduction band for this
structure is about $-0.15$.\cite{Pfeffer06} At the Fermi energy,
$E_F\simeq 25$~meV, its value is still less due to nonparabolicity
of the conduction band, it can be estimated as $\textsl{g}\simeq 0
\pm 0.05$. The results are presented in Fig.~\ref{F7}. It is seen
that even though $\sigma_{xx}$ changes from $45\,G_0$ down to
$5\,G_0$ in our magnetic field range [Fig.~\ref{F7}(a)], there is
no monotone change of the $K_{ee}$ value [Fig.~\ref{F7}(b)].
Exhibiting the weak SdH oscillations it remains constant. It
should be recorded that the value of $K_{ee}$ changes drastically
when the conductivity is lowered by means of the gate voltage: it
falls from $K_{ee}\simeq 0.4$ at $\sigma=50\,G_0$ down to
$K_{ee}\simeq 0.1$ at $\sigma=5\,G_0$.

Thus, the interaction contribution to the conductivity at $B=0$
decreases when the conductivity controlled by the gate voltage
goes down.  This decrease is well described by the RG equations
while $\sigma\gtrsim 15\,G_0$. On the other hand, no decrease of
the interaction contribution is evident when the conductivity
decreases being controlled by the external magnetic field. In this
case the contribution remains constant while $\sigma_{xx}$ is
lowered by one order of magnitude. Such the behavior remains
puzzling.

\section{Conclusion}
We have experimentally studied the evolution of the diffusion part
of the {\it e-e} interaction correction to the conductivity of 2D
electron gas in GaAs/In$_{0.2}$Ga$_{0.8}$As/GaAs single quantum
well within the conductivity range from $150\,G_0$ down to $\simeq
1\,G_0$.

To separate out the interaction contribution to the conductivity
we have used the unique property of the {\it e-e} interaction in
the diffusion regime, namely, the fact that it contributes to the
diagonal component of the conductivity tensor, $\sigma_{xx}$,
only. The simultaneous analysis of the weak localization
magnetoresistance, the temperature dependence of $\sigma_{xx}$ at
$B\neq 0$ and $\sigma$ at $B=0$ allows us to determine the
conductivity dependence of the Landau's Fermi liquid amplitude
$\gamma_2$. The results have been interpreted within the framework
of the RG theory. It has been obtained that the low temperature
value of  $\gamma_2$ increases with conductivity lowering. The
one-loop approximation of the RG theory adequately describes the
data while the conductivity is higher than $\simeq 15\,G_0$. At
lower conductivity, drastic disagreement between theory and
experiment is evident, suggesting the next-loop approximations in
the RG theory are needed. Finally, it remains unclear  why the
renormalization of the Fermi liquid amplitude takes experimentally
place when the conductivity decreases with the gate voltage, but
do not when it decreases in the external magnetic field. More work
is required to resolve this issue.

\subsection*{Acknowledgments}
We would like to thank I.~S. Burmistrov, I.~V. Gornyi,  and M.~V.
Sadovskii for illuminating discussions. This work was supported in
part by the RFBR (Grant Nos. 07-02-00528, 08-02-00662, and
09-02-0789).


\begin{thebibliography}{}
\bibitem{fnt2} By and large the structures 4261 and 3510 behave
    themselves analogously. Exception is that the slope of the
    $\sigma$~vs~$\ln{T}$ dependence is somewhat different from
    that predicted theoretically, $1+K_{ee}$. This issue will be
    discussed elsewhere.
\bibitem{fnt1}  We  have  supposed that the change
    of the slope of $\sigma$~vs~$\ln{T}$ and $\sigma_{xx}$~vs~$\ln{T}$
    dependences results from the renormalization of the interaction constant only. It is
    not straightforward method for the determination of the
    interaction constant. More direct method is based on the
    suppression of the triplet channel in the magnetic field due
    to the Zeeman effect. This method is applied for the systems
    with the large enough $\textsl{g}$-factor. It was used for the 2D hole system\cite{Minkov07}
    where $\textsl{g}\simeq 1.5-2$. However, the holes are the
    carriers of four-fold degenerated band with the total moment $3/2$.
    To the best of our knowledge,
    this method was not applied for the 2D electron
    systems with the simple, single-valley energy spectrum.
\end{thebibliography}
\end{document}